\documentclass[article]{aa} % for an article version
\usepackage{natbib}         % pour A&amp;A
\usepackage{graphicx}
\usepackage{longtable}
\usepackage[switch]{lineno} % switch=Put the line into the inner margin
\bibpunct{(}{)}{;}{a}{}{,} % to follow the A&amp;A style
\usepackage{amsmath}
\usepackage{subfig}
\usepackage{xcolor}
\newcommand{\ind}[1]{_{\mathrm{#1}}}
\newcommand{\diff}{\mathrm{d}}

\def\Kepler{\emph{Kepler}}

\def\numax{\nu\ind{max}}\def\nmax{n\ind{max}}

\def\Dnu{\Delta\nu}
\def\dnumoy{\langle\Delta\nu\rangle}

\def\Dnuas{\Delta\nu\ind{as}}
\def\epsas{\varepsilon\ind{as}}
\def\amp{{\mathcal{A}}}
\def\per{{\mathcal{G}}}
\def\phase{\phi}

\def\Msol{M_{\odot}}

\def\sup{\delta_{n}}
\def\supaj{\delta\ind{g,obs}}
\def\deltanuas{\Dnu_{\ind{UP}}(n)}
\def\deltanuobs{\Dnu(n)}

\def\ymodele{y_{i}(\textbf{a})}
\def\ydonnees{y_{i}}
\def\erreur{\sigma_{i}}
\def\Dnu{\Delta\nu}

\begin{document}
\title{Helium signature in red giant oscillation patterns observed by Kepler}
\titlerunning{Helium signature in red giants}
\authorrunning{Vrard, M et al.}
\author{Vrard M.\inst{1}, Mosser B.\inst{1}, Barban C.\inst{1}, Belkacem K.\inst{1}, Elsworth Y.\inst{2,} \inst{3}, Kallinger T.\inst{4}, Hekker S.\inst{5,} \inst{3,} \inst{6}, Samadi R.\inst{1}, and Beck P.G.\inst{7}
% rang 1
} \offprints{M. Vrard}

\institute{LESIA, CNRS, Universit\'e Pierre et Marie Curie,
Universit\'e Denis Diderot, Observatoire de Paris, 92195 Meudon cedex, France; \email{mathieu.vrard@obspm.fr}
\and
School of Physics and Astronomy, University of Birmingham, Edgbaston, Birmingham B15 2TT, UK
\and
Stellar Astrophysics Centre (SAC), Department of Physics and Astronomy, Aarhus University, Ny Munkegade 120, DK-8000 Aarhus C, Denmark
\and
Institute for Astronomy(IfA), University of Vienna, T\"urkenschanzstrasse 17, 1180 Vienna, Austria
\and
Max-Planck-Institut f\"ur Sonnensystemforschung, Justus-von-Liebig-Weg 3, D-37077 G\"ottingen, Germany
\and
Astronomical Institute “Anton Pannekoek,” University of Amsterdam, Science Park 904, 1098 XH Amsterdam, The Netherlands
\and
Laboratoire AIM, CEA/DSM-CNRS – Université Denis Diderot-IRFU/SAp, 91191 Gif-sur-Yvette Cedex, France
}
%\date{Written on}

\abstract{The space-borne missions CoRoT and $\Kepler$  have provided a large amount of precise photometric data. Among the stars observed, red giants show a rich oscillation pattern that allows their precise characterization. Long-duration observations allow for investigating the fine structure of this oscillation pattern}
{A common pattern of oscillation frequency was observed in red giant stars, which corresponds to the second-order development of the asymptotic theory. This pattern, called the universal red giant oscillation pattern, describes the frequencies of stellar acoustic modes. We aim to investigate the deviations observed from this universal pattern, thereby characterizing them in terms of the location of the second ionization zone of helium. We also show how this seismic signature depends on stellar evolution.}
{We measured the frequencies of radial modes with a maximum likelihood estimator method, then we identified a modulation corresponding to the departure from the universal oscillation pattern.}
{We identify the modulation component of the radial mode frequency spacings in more than five hundred red giants. The variation in the modulation that we observe at different evolutionary states brings new constraints on the interior models for these stars. We also derive an updated form of the universal pattern that accounts for the modulation and provides highly precise radial frequencies.}
{}


\maketitle

\voffset = 1.2cm
%________________________________________________________________

\section{Introduction\label{introduction}}

%\subsection{Context}

The space-borne photometric missions CoRoT \citep{2006ESASP1306...33B} and $\Kepler$ \citep{2010Sci...327..977B} have observed many red giants and led to substantial results. It has been shown that oscillations in red giants are similar to those seen in main-sequence solar-like stars \citep{2009Natur.459..398D,2010ApJ...713L.176B}. Characterization of the oscillation modes in red-giant spectra leads to reliable estimations of the mass and the radius of the stars \citep[e.g.,][]{2010A&A...522A...1K}. For many of them, it is possible to characterize the stellar core, hence provide information on the evolutionary state of the star \citep{2011Sci...332..205B,2011Natur.471..608B} and on the core rotation \citep{2012Natur.481...55B,2012A&A...548A..10M}.

Pulsating red-giant stars are characterized by two distinct resonant cavities, the core and the envelope. Gravity(g) waves only propagate in the radiative core. The inner turning point of the acoustic(p) modes is located near the external edge of the radiative region, and their outer turning point is near the surface of the star. In a red-giant star, the transition region between the two cavities is narrow, which leads to possibly efficient coupling between p and g waves and gives rise to the so-called mixed modes \citep[e.g.,][]{2011Sci...332..205B,2012A&A...540A.143M}. These modes behave as acoustic modes in the envelope and gravity modes in the core. Because of the nature of g-modes, the radial modes cannot couple, but dipolar and quadrupolar modes can be strongly coupled.

Radial solar-like oscillations, as observed in red giant spectra, have been identified as pressure modes that result from acoustic waves stochastically excited by convection in the outer layers of the star. The observed pressure mode pattern has been depicted in a canonical form, called the universal red-giant oscillation pattern \citep{2011A&A...525L...9M}.  This canonical form describes the regularity of the pressure mode pattern characterized by two quantities: the frequency $\numax$ of maximum oscillation and the mean frequency difference $\Dnu$ between consecutive pressure modes of same angular degree. From \citet{2013A&A...550A.126M}, we can consider the observed pattern to be the translation of the second-order asymptotic pattern described in \citet{1980ApJS...43..469T} at low radial order.
\newline

It has long been predicted that sound-speed discontinuities exist in star interiors and that they affect the observed solar-like oscillations \citep[e.g.,][]{1990LNP...367..283G}. In red giants, the main source of discontinuity is the region where helium undergoes its second ionization, as shown by \citet{2010A&A...520L...6M}. The sharp sound-speed variations produce a modulation in the observed oscillations frequencies, which is called a glitch. In this paper, we aim to investigate the deviations from the universal pattern, which are identified as glitches. Glitches related to the second helium ionization zone were first measured in the Sun by \citet{1991MNRAS.249..602D}. Their existence in other solar-like pulsators was also proposed \citep{1998IAUS..185..317M} and measured for the first time in a red giant by \citet{2010A&A...520L...6M}.

Glitches have been also observed in main-sequence stars showing solar-like oscillations \citep[e.g.,][]{2012A&A...540A..31M}. The study of glitches in red giants is intended to improve the understanding of their internal structure, to evaluate the amount of helium present in their envelopes and to characterize their evolutionary states \citep[e.g.,][]{2014MNRAS.tmp..576B}. For example, the amplitude of the glitch related to the region of second ionization of helium is directly correlated to the amount of helium present in the envelope.
\newline

We use stars with an evolutionary state already determined by \citet{2014arXiv1411.1082M} in order to investigate the behaviour of the glitches as a function of the evolutionary status. In Section 2, we detail the extraction of the radial frequencies of the stars. Section 3 provides a description of the oscillation pattern used as a global reference. In Section 4, we characterize the remaining modulation identified as glitches. In Section 5, we discuss the relationship between the measured parameters and compare them to predictions made by previous models \citep[e.g.,][]{2014MNRAS.tmp..576B,2014MNRAS.445.3685C}. Section 6 is devoted to conclusions.

\section{Data analysis method}

\subsection{Data set\label{Data.}}

We used the long-cadence data from $\Kepler$ with the maximum available length, up to the quarter Q17, corresponding to 1470 days of photometric observation. Original light curves were processed and corrected from phenomenological effects such as outliers, jump and drifts according to the method of \citet{2011MNRAS.414L...6G}. We used the set of 1142 stars for which \citet{2014arXiv1411.1082M} deduced the evolutionary status from the identification of the mixed-mode pattern. This allowed us to select oscillation spectra with a high signal-to-noise ratio as measured by an autocorrelation amplitude above $50$ \citep{2009A&A...508..877M}

\subsection{Radial mode fitting method\label{Radial_mode.}}

As indicated previously, dipolar and quadrupolar modes in red giant stars oscillation spectra have a mixed component. They have, then, a complex frequency pattern. Consequently, we have focused on radial modes only.
\newline

The universal pattern predicts the positions of modes using the so-called large separation which is the average frequency separation between modes of the same degree and consecutive order. A first estimate of the frequency position of radial modes is determined with this universal pattern \citep{2011A&A...525L...9M}. This guess is refined by automatically locating the nearby local maxima in the smoothed spectrum within a range of frequencies equivalent to a tenth of the large frequency separation. These local peaks are considered to be reliable modes when their heights are above a threshold value corresponding to the rejection of the H$_{0}$ hypothesis with a confidence level of 99.9$\%$.

\begin{figure}                 % Insertion d'une figure = objet flottant
  \includegraphics[width=9cm]{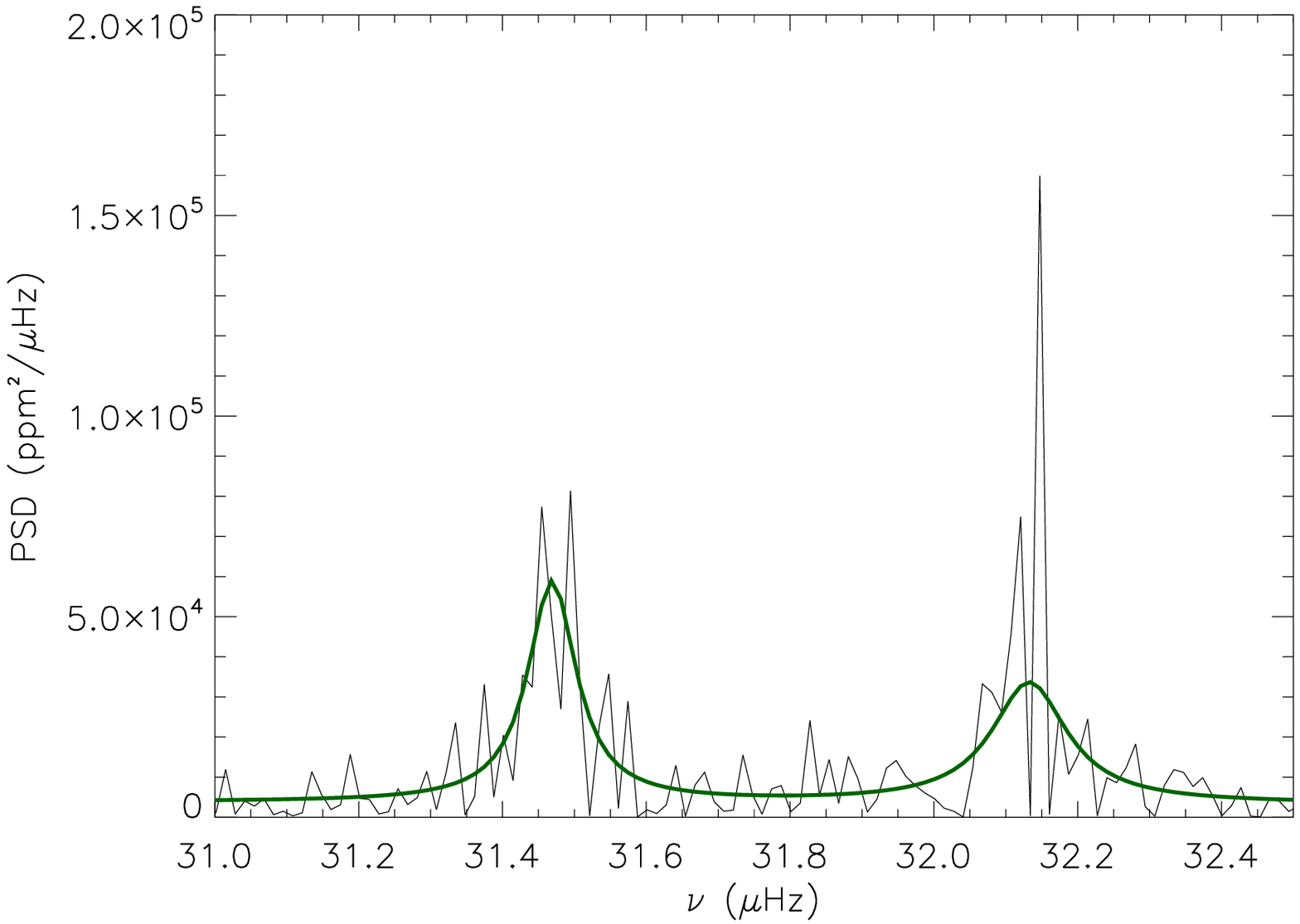}
  \caption{Power density spectrum (PSD) of the red clump star} KIC2303367. The peak around 32.1 $\mu$Hz is a radial mode and the peak on the left is the nearby $\ell=2$ mode. The Lorentzian fits are shown with the solid green line.
  \label{fig:lorentzienne}
\end{figure}

We then consider small frequency windows around each separate radial mode.
We fit a Lorentzian model to these modes using the maximum likelihood estimator technique described in \citet{1994A&A...289..649T} and \citet{2014aste.book..123A}. The structure is fitted as a number of Lorentzians plus a background component (Fig.~\ref{fig:lorentzienne}) following the model of \citet{2010AN....331.1016B}

%Ecrire P(v) auparavant puisqu'on ne l'introduit pas.

\begin{equation}
   P(\nu) = \sum_{k=1}^{Q} M(k,\nu) + B(\nu),
   \label{Modelled_spectrum}
\end{equation}
where $Q$ is the number of modes modelled by a Lorentzian function $M(k,\nu)$, and $B(\nu)$ is the background noise in the power spectrum modelled as a linear function of $\nu$, i.e $a+b\nu_{k}$. Several alternative forms of the background were tried \citep[e.g.,][]{2011ApJ...741..119M}, and we found that a change in the background component has a negligible impact on the measured frequencies of the radial modes. $Q$ may be greater than 2 when dipole mixed modes are present or when the quadrupole modes are split by rotation.

$M(k,\nu)$ is a Lorentzian profile

\begin{equation}
   M(k,\nu) = \frac{H_{k}}{1+\left(\frac{2(\nu-\nu_{k})}{\Gamma_{k}}\right)^{2}},
   \label{Modelled_lorentzian}
\end{equation}
where $H_{k}$ is the height of the Lorentzian profile, $\nu_{k}$ is the oscillation mode frequency, and $\Gamma_{k}$ is the mode linewidth (FWHM) with $\Gamma_{k}=1/(\pi\tau)$, $\tau$ being the mode lifetime. This method provides realistic uncertainties on the determination of the mode frequency. The mean value of the uncertainties we obtained for our set of stars is about $10$ nHz. We checked that the modelling of the background does not modify the fit of the modes frequencies and found that it has a negligible impact on the frequency position measured for the fitted modes. The position of the radial modes are then checked by eye for verification purposes: a non-radial mode could be mistakenly identified as a radial mode.

\section{Glitches inference}

The aim of this section is to describe how we measure acoustic glitches. This measurement relies on the precise identification of radial-mode frequency spacing. Since there is no unique definition of those spacings, we have to introduce different notations. The mean observed large separation is denoted $\dnumoy$. A weighted mean large separation over a broad range of order was used to measure this parameter. The observed difference between individual radial mode frequencies is denoted by $\deltanuobs$ and $\deltanuas$ is used for the reference values provided by the universal pattern.

\subsection{Determination of the global large separation $\dnumoy$}

Since red giant stars pulsate at medium to low radial order (the typical radial orders at which red giant stars pulsate are about 10 by comparison with around 20 for main-sequence stars), second-order terms in the asymptotic expansion need to be considered.
\newline
%They are translated in a curvature in the \'echelle diagram.

The autocorrelation method described in \citet{2009A&A...508..877M} provides a first approximation of the global large separation $\dnumoy$ and of the frequency $\numax$ of maximum oscillation power. As with any method, this automated method is affected by the so-called realisation noise due to stochastic excitation of short-lived waves (Fig.~\ref{fig:lorentzienne}). A way to lower this noise is provided by the universal pattern \citep{2011A&A...525L...9M}. In practice, the correlation of the observed spectrum with a synthetic spectrum allows us to substantially reduce this noise. It follows that this automated method gives precise results, even at low frequency \citep{2012A&A...544A..90H}.

The universal pattern as introduced by \citet{2011A&A...525L...9M} for red giants describes the second-order term of the asymptotic expansion with a curvature term $\alpha$:

\begin{equation}
   \nu_{\ind{UP}}(n,0) = \left(n + \varepsilon + \frac{\alpha}{2}(n - \nmax)^{2}\right)\dnumoy ,
   \label{dnu_local_lisse}
\end{equation}
where $\varepsilon$ is the asymptotic offset and $\nmax$ is the radial order corresponding to the oscillation of maximum amplitude defined by $\nmax = \numax/\Dnu - \varepsilon$. Note that $\nmax$ is not an integer.

\subsection{Determination of the frequency differences $\deltanuobs$}

We calculate the local frequency separation by computing the frequency differences between consecutive radial modes. We derive the local $\deltanuobs$ from the average over adjacent modes,

\begin{equation}
   \deltanuobs = \frac{\nu_{n+1,0}-\nu_{n-1,0}}{2} ,
   \label{dnu_theorique}
\end{equation}
where $n$ is the radial order. At the edges of the measured radial modes, we cannot use this equation and replace it by the frequency difference between two consecutive radial modes. The frequency reference for each difference is taken as the mid-point as the data values used to derive the difference. The use of Eq. (\ref{dnu_theorique}) to derive the frequency separation introduces correlations since every radial mode frequency is used for estimating two values of $\Dnu(n)$. The frequency separation $\Dnu (n)$ is correlated with $\Dnu(n \pm 2)$ but uncorrelated with  $\Dnu(n \pm 1)$. This correlation is later taken into account in the error budget.

The results for one star are shown in the top panel of Fig.~\ref{fig:avant}. Two notable features can be noticed: an upward trend with increasing frequency which we associate with the second-order term in the asymptotic expansion, and a modulation that we later attribute to glitches.

\subsection{Curvature}

We can deduce the asymptotic dependence of the large separation with the frequency from the derivative of Eq. (\ref{dnu_local_lisse}) with respect to the radial order $n$:

\begin{equation}
    \deltanuas = (1+\alpha(n-\nmax))\dnumoy .
    \label{dnu_local_curvature}
\end{equation}
According to the universal pattern, the variation of $\deltanuas$ is linear and only depends on the curvature term. As this asymptotic development is only valid for smooth interiors, inner discontinuities will lead to a departure from that pattern. In order to identify any modulation due to internal discontinuities, we have to remove the curvature term.

\begin{figure}                 % Insertion d'une figure = objet flottant
  \includegraphics[width=9cm]{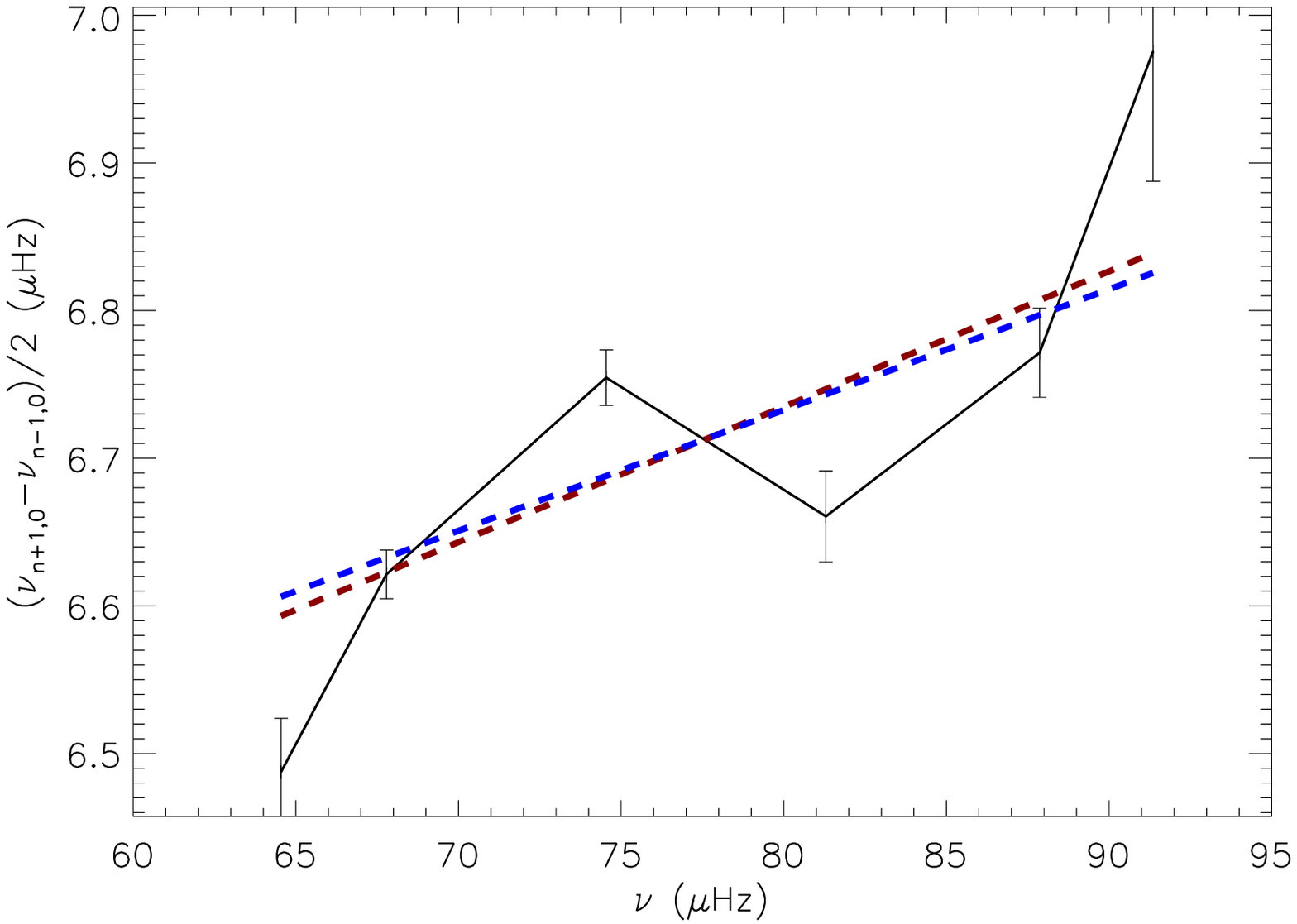}
  \includegraphics[width=9cm]{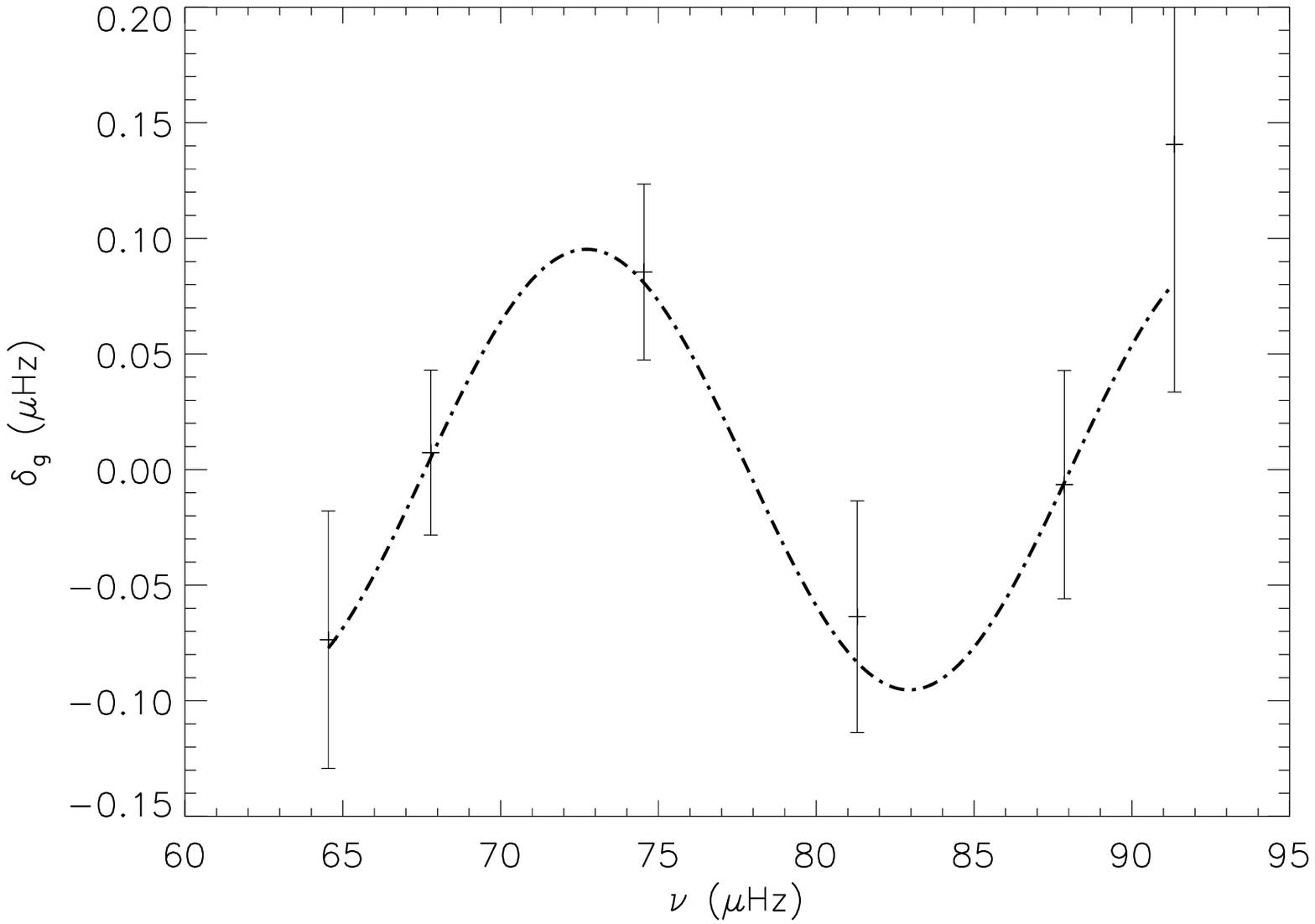}
  \caption{\emph{Top}: Variation of the large separation as a function of frequency for the star KIC1872166. Error bars correspond to the 1-$\sigma$ uncertainties. The general trend is due to the second-order term (the curvature) and the modulation is due to the presence of acoustic glitches. The red dashed line is a linear fit over the oscillations to estimate the curvature parameter. The blue dashed line is the value of the mean curvature given by \citet{2013A&A...550A.126M}
\emph{Bottom}: Variation of the large separation $\sup$ as a function of frequency after suppressing the curvature term. The dotted-dashed line shows a fit obtained for this star using Eq. (\ref{resultante}).}
  \label{fig:avant}
\end{figure}

We used the relation

\begin{figure}                 % Insertion d'une figure = objet flottant
  \includegraphics[width=9cm]{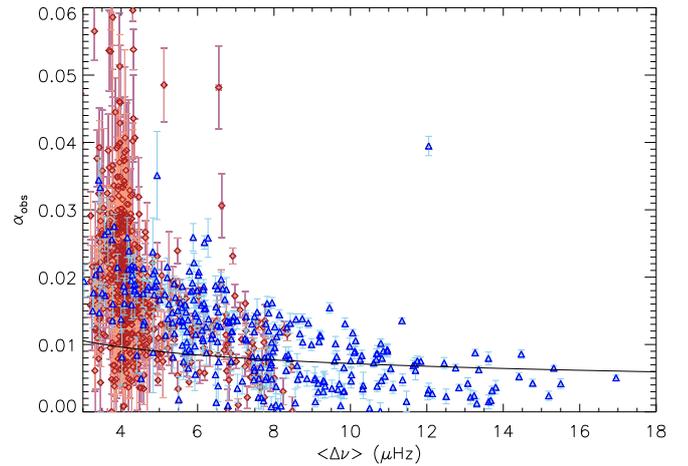}
  \caption{Apparent curvature $\alpha_{obs}$ as a function of the large separation for each red giant stars. Clump stars are indicated by red diamonds and RGB stars by blue triangles. Error bars correspond to the $1\sigma$ uncertainties. The solid black line correspond to the value of the mean curvature $\alpha$ found by \citet{2013A&A...550A.126M}}
  \label{fig:curvature}
\end{figure}

\begin{equation}
    \alpha = 0.015 \dnumoy^{-0.32} ,
    \label{alpha_measured}
\end{equation}
which provides a consistent glitch-free description of the curvature term averaged over a large number of stars \citep{2013A&A...550A.126M}. In fact, the curvature observed in an individual star is affected by the existence of glitches. This is illustrated in Fig. \ref{fig:avant} (top), where it is evident that the linear term is steeper when it is derived locally. Using it would yield an overestimate of the asymptotic second-order effect \citep{2013A&A...550A.126M}. \citet{2013MNRAS.434.1668H} showed that the curvature is significantly overestimated when only a small number of radial modes are used to derive it. We therefore use Eq. (\ref{alpha_measured}) as a reliable formulation to specify $\alpha$. We also tested if a change in the $\alpha$ parameter modifies the results obtained for the glitch measurement. We found that it is important to use a consistent value of $\alpha$ for all stars but the precise formulation is not crucial as long as the changes are small (see section 4.4).

\section{Measurements\label{Mesure}}

\subsection{Measuring the glitches for red giants}

The methods used for measuring the modulation due to the glitches were mostly developed for main-sequence stars \citep{1994A&A...283..247M,2005MNRAS.361.1187M,2007MNRAS.375..861H}. Two discontinuities are mainly considered: the base of the convection zone and the helium second ionization region. In red giants, the dominant discontinuity is the second helium ionization zone, as shown by \citet{2010A&A...520L...6M}. This work also pointed out that there are too few identified modes to take all discontinuities into account. Since the discontinuity of the second helium ionization is by far the most important one in red giants, we can assume that there is only one modulation component \citep[Fig 2. of][]{2010A&A...520L...6M}. Most of the methods used to characterize the glitches describe them in a complex way with a frequency-dependent amplitude \citep{2005MNRAS.361.1187M,2007MNRAS.375..861H,2014ApJ...782...18M}. Here, due to the low number of modes measured and the large uncertainties, we prefered to use that simple fit.
\newline

We compute the difference between the observed local large separation and the theoretical local large separation predicted by the universal pattern:

\begin{equation}\label{suppress}
    \supaj = \deltanuobs - \deltanuas .
\end{equation}

We fitted an oscillatory component to the resultant frequency variations obtained after removal of the curvature term from the measurements:

\begin{equation}\label{resultante}
     \supaj = \amp\dnumoy\cos\left(\frac{2\pi(\nu-\numax)}{\per\dnumoy} + \phase\right) ,
\end{equation}
where $\per$ is the period of the oscillation expressed in units of $\dnumoy$, $\amp$ is the amplitude of the oscillation in units of $\dnumoy$ and $\phase$ is the phase of the oscillation centered on $\numax$. We used a $\chi^{2}$ minimisation to fit the three free parameters. The uncertainties were extracted by the inversion of the Hessian matrix constructed with a covariance matrix (see Appendix \ref{ap:1}). This method takes the correlation of the frequency difference calculated with Eq. (\ref{dnu_theorique}) into account. An example of a fit is shown in the bottom panel of Fig. \ref{fig:avant}.

In some of the stars we analysed, the fit was unsuccessful. This happens, for example, when the uncertainties on the frequencies are too large. We choose to reject such fits when the uncertainties on the period parameter were higher than the measured value of the period.

It is important to note that the uncertainties on $\numax$ impact on the uncertainties measured for the phase parameter. The error on $\numax$ corresponds typically to a fifth of the $\dnumoy$ value. Therefore, following Eq. (\ref{resultante}), an error on $\numax$ will translate into an error on the phase parameter of about $2\pi/(5\per)$. With $\per$ $\sim4$, this will quadratically add an error of 0.3 rad on the phase. It is worth noting, that this value is much smaller than the pattern found in Section 4.4 for this parameter.

\subsection{Period of the modulation}

\begin{figure}                 % Insertion d'une figure = objet flottant
  \includegraphics[width=9cm]{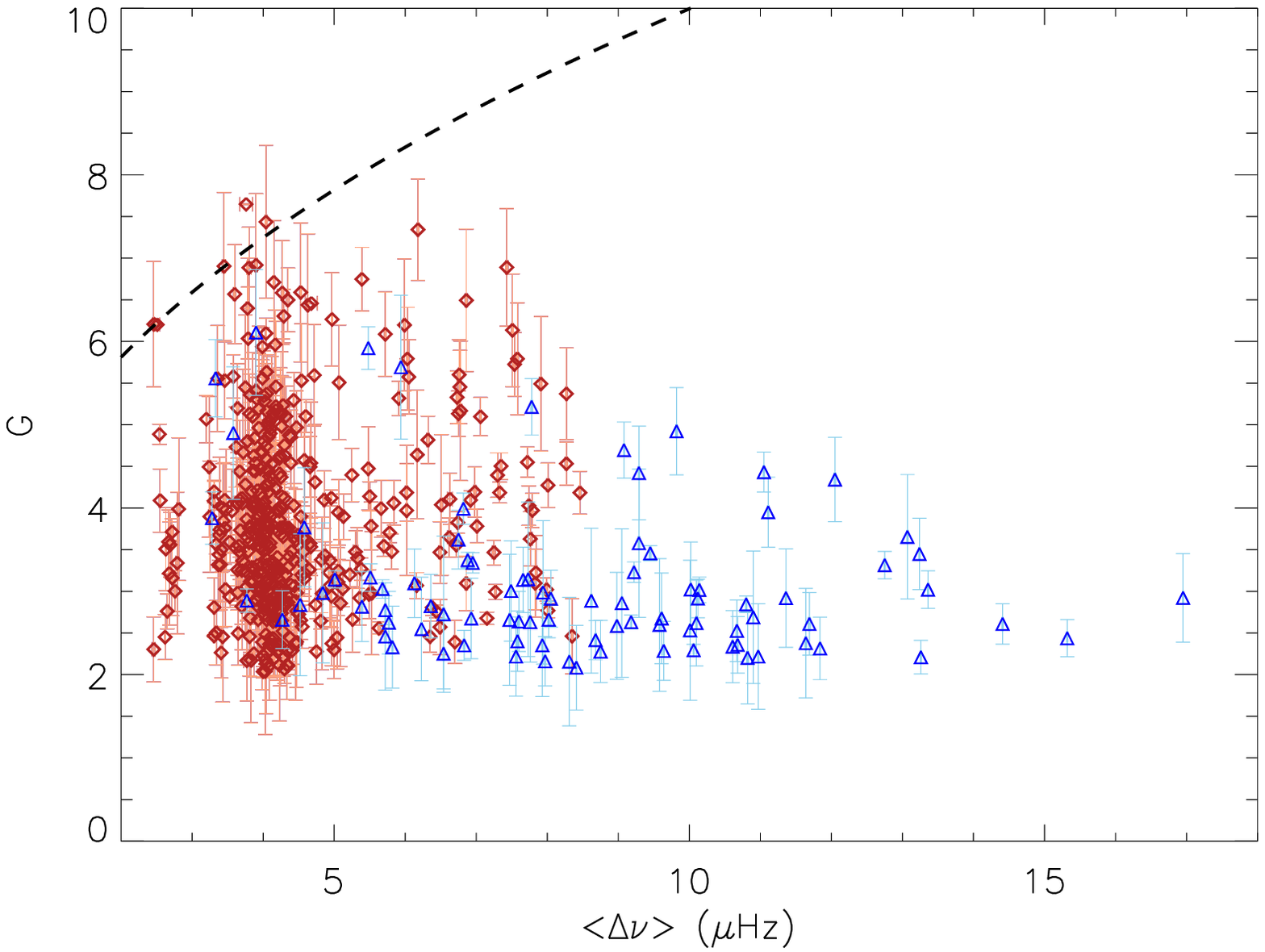}
  \includegraphics[width=9cm]{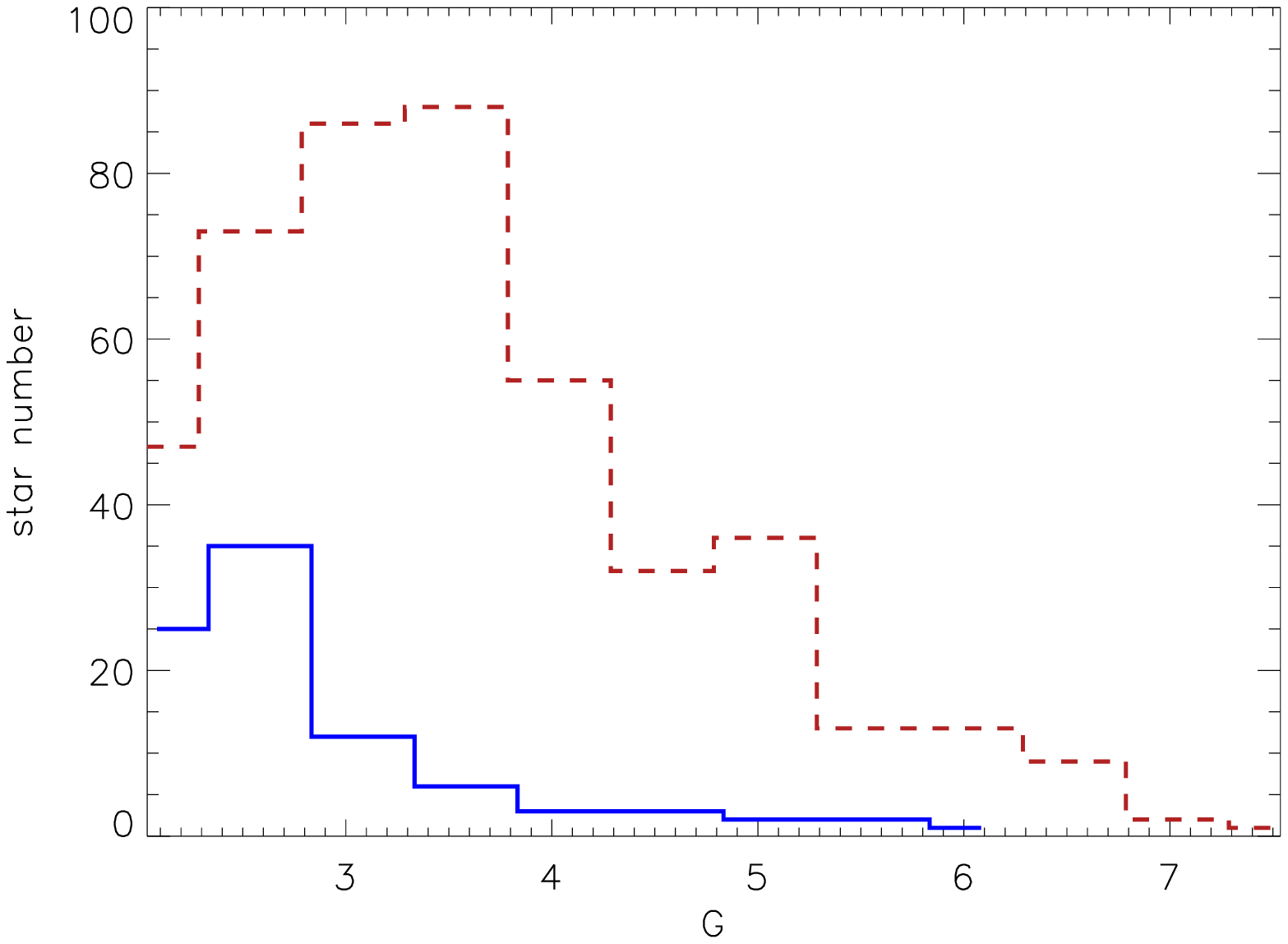}
  \caption{\emph{Top}: Dimensionless period $\per$ of the modulation measured as a function of the large separation. Clump stars are indicated by red diamonds and RGB stars by blue triangles. Error bars correspond to the $1\sigma$ uncertainties. The dashed black line indicates the maximum number of radial modes observable in a red giant spectrum.
\emph{Bottom}: Histogram of the dimensionless period $\per$. Clump stars are indicated by the red dashed line and RGB stars by the blue line.}
  \label{fig:period}
\end{figure}

We measured the dimensionless period $\per$ for 546 stars. The variation of the period as a function of the large separation is shown in Fig. \ref{fig:period}. The dimensionless period $\per$ shows no significant trend with $\dnumoy$ over a large range. The clump star values show a higher dispersion than the RGB stars. We can however note a second branch of RGB stars with much higher period values. As of now, there is no explanation toward this behaviour. The mean glitch periods are $\per\simeq3.08\pm0.65$ for RGB and $\per\simeq3.83\pm0.88$ for clump stars, higher than for RGB stars by approximately 30$\,\%$. Such values are similar to the periods predicted by the models for this kind of star \citep{2014MNRAS.tmp..576B}.

The values found for the glitch periods are less than the typical number of radial modes measured in red giant, which is around 6. As the global large separation measurement takes all radial modes frequencies into account, we infer that glitches are smoothed out when the mean large separation is measured globally. As a consequence, the glitches have limited influence on a global measurement of the large separation. On the contrary, they may considerably affect local measurements, as shown by \citet{2012A&A...541A..51K}.
\newline

\subsection{Amplitude of the modulation}

\begin{figure}                 % Insertion d'une figure = objet flottant
  \includegraphics[width=9cm]{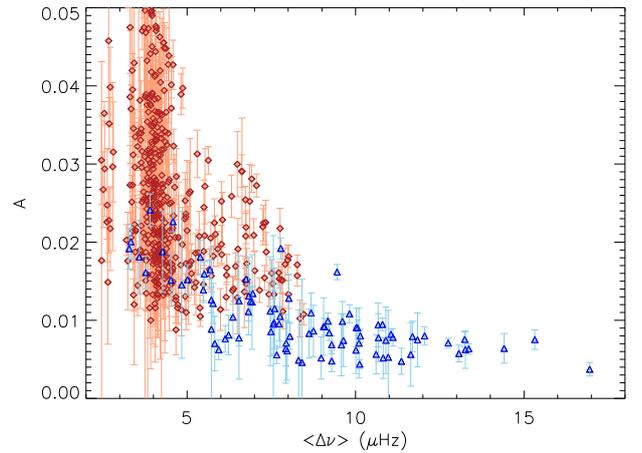}
  \caption{Dimensionless amplitude $\amp$ of the modulation measured as a function of the large separation. Clump stars are indicated by red diamonds and RGB stars by blue triangles. Error bars correspond to the $1\sigma$ uncertainties.}
  \label{fig:amplitude}
\end{figure}

%\begin{figure*}                 % Insertion d'une figure = objet flottant
%  \includegraphics[width=18cm,height=7cm]{result_glitches_ligne.eps}
%  \includegraphics[width=18cm,height=7cm]{result_amplitude_pres.eps}
%  \includegraphics[width=18cm,height=7cm]{result_phase_bien.eps}
%  \caption{a: Dimensionless period $\per$ of the modulation measured as a function of the large separation. Clump stars are indicated by red diamonds and RGB stars by blue triangles. Error bars correspond to the $1\sigma$ uncertainties. The dashed black line indicates the maximum number of radial modes observable in a red giant spectrum.
%b: Dimensionless amplitude $\amp$ of the modulation measured as a function of the large separation. Clump stars are indicated by red diamonds and RGB stars by blue triangles. Error bars correspond to the $1\sigma$ uncertainties.
%c: Phase $\phase$ of the modulation measured as a function of the global large separation. Clump stars are indicated by red diamonds and RGB stars by blue triangles. Error bars correspond to the $1\sigma$ uncertainties.
%}
%  \label{fig:amplitude}
%\end{figure*}

The variation of the dimensionless amplitude $\amp$ as a function of the large separation is plotted in Fig. \ref{fig:amplitude}. It shows that the relative amplitude of the modulation $\amp$ for RGB stars increases when the large separation decreases as a consequence of stellar evolution. Clump stars have roughly consistent amplitudes within the uncertainties. There is some evidence that the amplitudes for secondary clump stars are higher than for RGB stars. We checked that these phenomena were not due to the exclusion of a frequency dependent amplitude in the fit.

We can now derive a relationship between the relative amplitude and $\dnumoy$.

%$a = 0.07\pm0.01$ and $b = -0.74\pm0.05$ for clump stars, and 
\begin{equation}\label{amplitude_variations}
    \amp = a\dnumoy^{b} ,
\end{equation}
with $a = 0.06\pm0.01$ and $b = -0.88\pm0.05$ for RGB stars with $\dnumoy$ in $\mu$Hz. The absolute amplitude (i.e. the product of $\amp$ and $\dnumoy$) shows almost no dependence with $\dnumoy$ since the power index $b$ is close to $-1$, as derived from models \citep{2014MNRAS.tmp..576B}.

Stellar evolution models show that the amplitude of the discontinuity in the adiabatic exponent increases when the star evolves up the RGB. Such an increase is also present in the sound-speed profile \citep[e.g.,][]{2011MNRAS.418L.119H}. The amplitude of the discontinuity strongly depends on the amount of helium in the convective envelope of the star. During the red giant evolution, a structure change at, e.g., the luminosity bump could bring additional helium in the convective envelope of the star. As a result, a larger discontinuity due to the helium second ionization zone is expected. It could explain why clump stars have a slightly larger observed glitch amplitude than RGB stars below the bump.

\subsection{Phase of the modulation}

\begin{figure}                 % Insertion d'une figure = objet flottant
  \includegraphics[width=9cm]{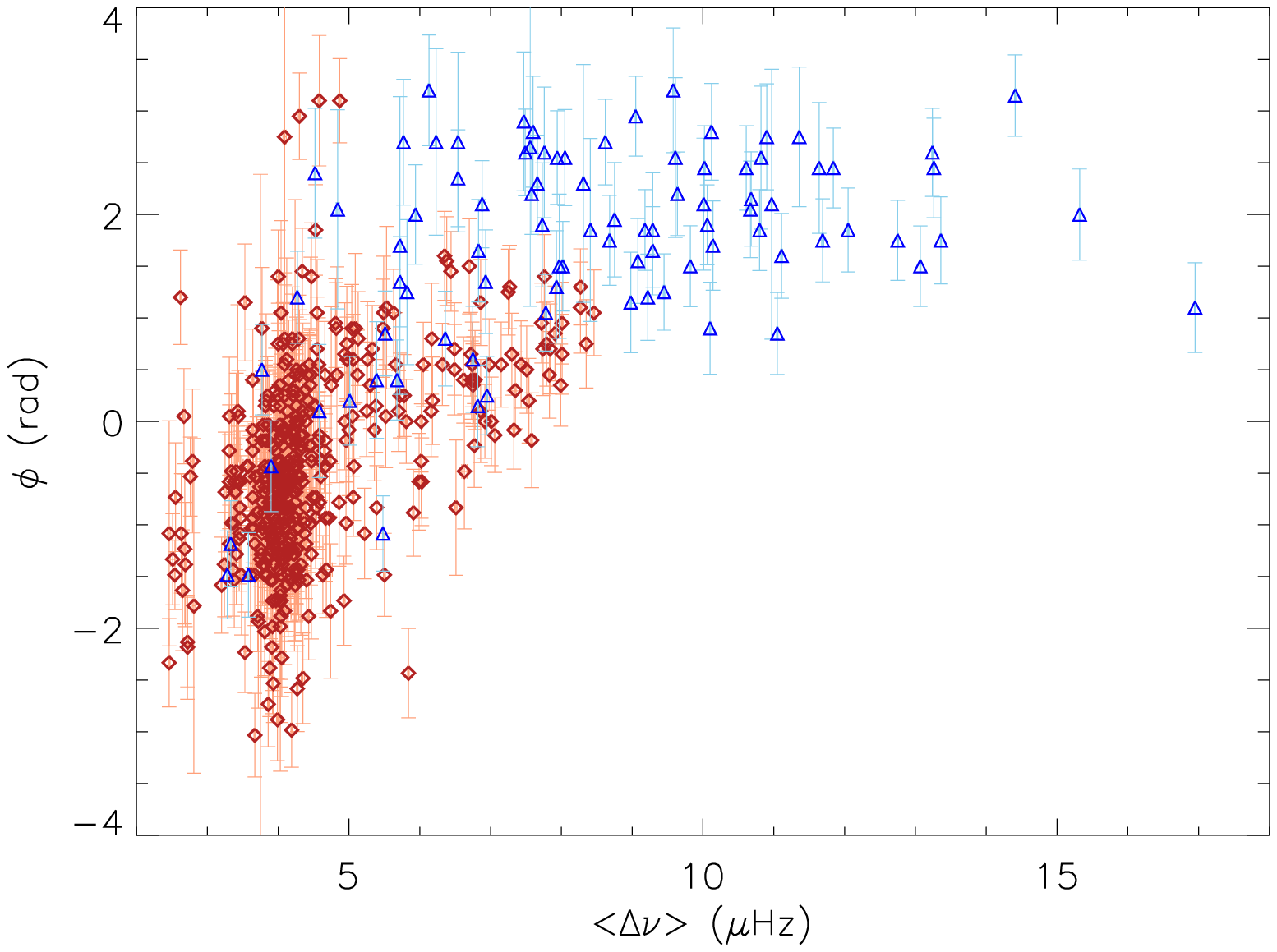}
  \includegraphics[width=9cm]{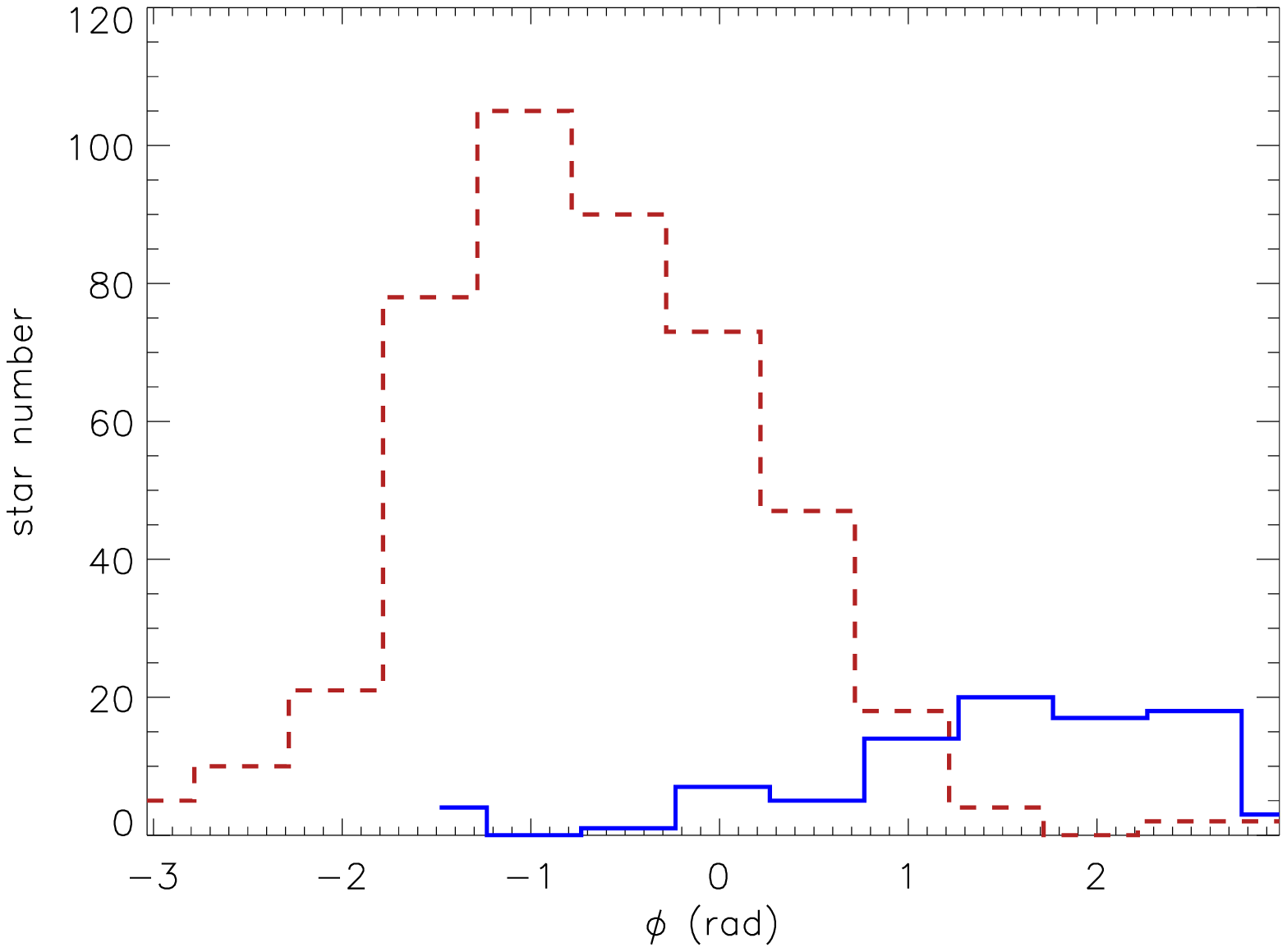}
  \caption{Phase $\phase$ of the modulation measured as a function of the global large separation. Clump stars are indicated by red diamonds and RGB stars by blue triangles. Error bars correspond to the $1\sigma$ uncertainties.
\emph{Bottom}: Histogram of the phase $\phase$. Clump stars are indicated by the red dashed line and RGB stars by the blue line.}
  \label{fig:phase}
\end{figure}

We now turn to a consideration of the measured phase $\phi$ of the oscillation due to the glitch. This parameter shows complex variation (Fig.~\ref{fig:phase}). All clump stars have a phase between $-2$ rad and $1$ rad, whereas RGB stars have $\phi$ between $1$ rad and $3$ rad. This phase shift between clump and RGB stars is systematically observed. We investigated the influence of the curvature term on these results and found that a curvature change as high as $50\, \%$ does not influence the trend. Thus, the difference between clump and RGB stars is real and is not an artefact of the data processing.
\newline

To investigate the consequences of this phase difference, we consider the phase at the order corresponding to the index $\nmax$ of the maximum oscillation signal. Following Eq. (\ref{resultante}), if the star has a phase $\simeq$ 0 like clump stars, the local large separation measured will be overestimated. On the contrary, if the star has a phase $\simeq$ $\pi$ like RGB stars, the local large separation will be slightly underestimated, as shown in Section 5.5. This property can be used to determine the evolutionary stage of the stars, as discussed in the section 5.3.

Additionally, we note the transition to lower values of the $\phi$ parameter for RGB stars with a large separation below $6\,\mu$Hz. This could possibly correspond to the signature of a variation in internal structure related to the luminosity bump \citep{2012A&A...543A.108L}.

\subsection{On simulated data}

We have performed tests to evaluate the statistical stability of the fit of the glitches (Eq. \ref{resultante}). First, we have tested the influence of pure noise on the universal oscillation pattern. Synthetic radial mode frequencies have been created, based on the asymptotic pattern and including a sinus-like modulation. A random shift has been added to each frequency, corresponding to a gaussian noise component. We then deduced the local large separation from these frequencies. We have used the fitting program to extract the components of a spurious glitch signal introduced by this noise component. The results obtained with one thousand simulations indicate that there is a preferance in favour of the detection of spurious, low period signals ($\per$ around 2.5) independent of the large separation or the signal-to-noise ratio. These low periods are associated with random phases and amplitudes. These characteristics are typical of the white noise and do not correspond to the signal we observe. Hence, this first set of simulation proves that what we observe for the phase and period parameters is definitely not due to noise.

We performed a second test, with the addition of a sinusoidal modulation to the oscillation pattern. Various levels of noise were considered. The parameters of the sinusoidal component could be retrieved, even at low signal-to-noise ratio ($S/N = 1$). These results confirmed that the analysis is relevant, even when only five radial orders are considered. We however noticed a small bias on the inferred period. Further tests have been performed to provide an estimate of this bias. For a modulation period $\per$ of the sinusoidal component, the fit provides a period $\per\star = 1.03\per + 0.64$. Periods were calibrated to take that phenomenon into account. Similar tests have shown that the phases and amplitudes are reliably measured. We also tested for the importance of the number of orders included in the data. We found that a modification from $10$ to $5$ radial orders give consistent results within $10$ $\%$.

\section{Discussion\label{Discussion_bis}}

In this section we interpret our results in terms of physical quantities and compare them to stellar models. The modelling of the glitch is then combined with the universal pattern to provide a precise fit of radial frequencies.

\subsection{Acoustic radius of the HeII ionization region}

Any structure discontinuity induces a seismic signature characterized by its acoustic radius \citep[e.g.,][]{1990LNP...367..283G}. The measured period $\per$ is directly related to the acoustic depth of the glitch \citep{2014ApJ...782...18M}

\begin{equation}\label{profondeur acoustique}
    \frac{1}{2\per\dnumoy} = \tau_g = \int_{r_g}^{R_s} \frac{\diff r}{c_s} ,
\end{equation}
where $c_s$ is the adiabatic sound speed, $\tau_g$ is the acoustic depth of the glitch, $r_g$ is its radial position and $R_s$ is the seismic radius of the star. If the centre of the star is taken as a reference, we can define the acoustic radius as:

\begin{equation}\label{rayon_acoustique}
    T_g = \int_{0}^{r_g} \frac{\diff r}{c_s} .
\end{equation}
We can relate the two physical parameters by the relation $\tau_g = T_0 - T_g$, with $T_0$ the total acoustic radius of the star. The large separation is also related to the stellar acoustic diameter. However, the global large separation $\dnumoy$ we measure is different from the theoretical large separation, called asymptotic large separation $\Dnuas$ \citep{2013A&A...550A.126M}

\begin{equation}\label{deltanu}
    \Dnuas = \left(2\int_{0}^{R} \frac{\diff r}{c_s}\right)^{-1} ,
\end{equation}
where $c_s$ is the sound speed. However, \citet{2013ASPC..479...61B} have shown that $\dnumoy$ provides a valuable approximation for deducing the mass up to $10\,\%$ and radius of the star up to $5\,\%$. Hence, we use it in this work.

Since the large separation depends on the total acoustic radius of the star, we can write $T_0 = 1/(2\dnumoy)$. Thus, we can relate the acoustic radius to the period measured:

\begin{figure}                 % Insertion d'une figure = objet flottant
  \includegraphics[width=9cm]{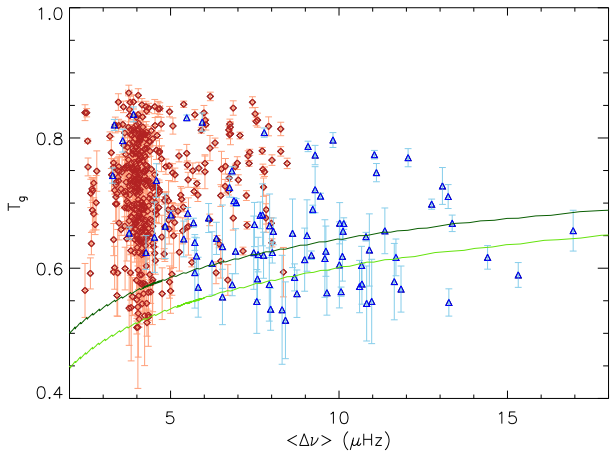}
  \caption{Acoustic radius of the discontinuity related to the second helium ionization zone as a function of the global large separation. Clump stars are indicated by red diamonds and RGB stars by blue triangles. Error bars correspond to the $1\sigma$ uncertainties. The light green line indicates the theoretical acoustic radius of the second helium ionization zone for a $1 \Msol$ star during the RGB phase. The dark green line gives the same information for a $1.4 \Msol$ star.}
  \label{fig:profondeur}
\end{figure}

\begin{equation}\label{rayon_acoustique1}
    T_g = \frac{1}{2\dnumoy}\left(1-\frac{1}{2\per}\right) .
\end{equation}
The measured $T_g$ are compared to results from CESAM2k models in Fig.~\ref{fig:profondeur}. The models used are described in \citet{2012A&A...540L...7B}. The conclusions obtained for these models are confirmed by other simulations done by, e.g., \citet{2010A&A...520L...6M}.
\newline

There is a reasonable agreement between the observed and modelled values of the acoustic radius for RGB stars but a clear discrepancy for clump stars (Fig. \ref{fig:profondeur}). Simulations show that a clump star and a RGB star with the same mass and radius should have about the same acoustic radius $T_g$. This is not what we observe. These are the first extensive masurements of the acoustic radius of the helium ionization zone in clump stars, the discrepancy that we observe could either be due to the difficulty to model clump stars, or to another poorly known phenomenon like the mass loss at the tip of the red giant branch. We leave an exploration of the reasons for this discrepancy to future work.

\subsection{Measuring the helium abundance with glitches?}

\citet{2014MNRAS.tmp..576B} have shown that red giants with similar $\dnumoy$ but with different amount of helium show very similar glitch amplitude. Owing to the large uncertainties we observe on these amplitudes, we have not been able to extract information on the helium abundance present in these stars. Furthermore, at this stage we do not identify any way to improve the measurement and derive quantitative information on the helium content in red giants from the glitches.

\subsection{Measuring the evolutionary stage}

Previous work has shown a partition of red giants, depending on their evolutionary states, based on the $\varepsilon$ offset \citep[Fig. 3b of][]{2011Natur.471..608B,2012A&A...541A..51K}. This parameter $\varepsilon$ in the asymptotic expansion is equal to $1/4$ in the asymptotic relation \citep{1980ApJS...43..469T}. Its value transferred in the universal pattern (Eq. \ref{dnu_local_lisse}) is significantly different and varies with $\dnumoy$. The link between $\epsas=1/4$ and the observed values $\varepsilon(\dnumoy)$ is however totally explained by the use of different definitions of the large separation \citep{2013A&A...550A.126M}. The difference $\Delta\varepsilon = \varepsilon\ind{clump}-\varepsilon\ind{RGB}$ between clump stars and RGB stars with similar $\dnumoy$ is, according to \citet{2012A&A...541A..51K}, $-0.15$. These results are derived from a local analysis of the radial spectrum.

Our approach, based on a global description with a global measurement of the large separation and a generic description of the radial oscillation pattern, gives an interpretation of the physical origin of the difference. We did not find any $\varepsilon$ offset between clump stars and RGB stars but we observed that the glitch component modifies the local measurement of the large separation with a relative variation of $-0.5\,\%$ for RGB stars and $+1\,\%$ for clump stars. This translates into a change in the $\varepsilon$ parameter ($\delta\varepsilon = -(n+\varepsilon)\,\delta \log(\Dnu)$ according to the derivation of Eq. (\ref{dnu_local_lisse})), corresponding to  $\varepsilon\ind{clump}=+0.05$ for RGB stars and $\varepsilon\ind{RGB}=-0.1$ for clump stars with $\dnumoy = 4 \mu$Hz. The difference between clump stars and RGB stars is therefore $\Delta\varepsilon=-0.15$ in agreement with local measurements \citep{2012A&A...541A..51K}.
 
The difference reported by \citet{2012A&A...541A..51K} has been noted for a vast majority of red giants. We can therefore conclude that our results, reduced to a limited subset of red giants showing enough oscillation modes, can be extended to all red giants. Therefore, this work justifies the analysis made by \citet{2012A&A...541A..51K} for distinguishing RGB and clump stars. Alternatively, measuring the phase shift (Eq. \ref{resultante}) provides similar information.

So, we have now two methods for measuring the evolutionary stage of red giants, either based on the mixed modes \citep[e.g.,][]{2011Natur.471..608B,2012A&A...540A.143M}, or based on the glitch phases \citep{2012A&A...541A..51K}. Grosjean et al. (2014) recently confirmed that gravity-dominated mixed modes in low-mass stars with $\Dnu$ less than $6\,\mu$Hz have small heights. This hampers the use of mixed modes for determining the evolutionary stage and reinforces the importance of the measurement of the evolutionary states only from radial modes.
\newline

\subsection{Mass dependence}

\begin{figure}                 % Insertion d'une figure = objet flottant
  \includegraphics[width=9cm]{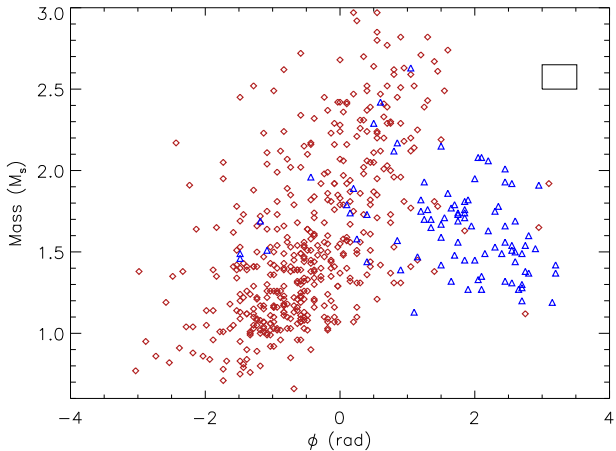}
  \caption{Stellar masses in function of the phase of the modulation. Clump stars are indicated by red diamonds and RGB stars by blue triangles. 
The black box in the right corner of the figure corresponds to the typical $1\sigma$ uncertainties for these parameters
}
  \label{fig:phase_mass}
\end{figure}

Based on the measured $\dnumoy$ and $\numax$ and on the temperature of the star given by \citet{2014ApJS..211....2H}, we deduced the approximate mass and radius of the star from the usual scaling relations \citep{2010A&A...522A...1K,2013A&A...550A.126M}. We then investigated the mass dependence of the different glitch modulation parameters. Only the phase shows a clear variation with the stellar mass (Fig. \ref{fig:phase_mass}). We did not find any correlations between the mass and other parameters (period and amplitude). Phases of the clump stars have a clear mass dependence: clump stars with a high mass have a higher phase than clump stars with a lower mass. This mass dependence is not seen for RGB stars.

The observed relation between the phase and the mass is clear but remains purely empirical. Its physical basis needs to be established. More theoretical work is therefore needed, as a follow up to, e.g., \citet{1993A&A...274..595P,2014MNRAS.445.3685C} for p modes and \citet{2008MNRAS.386.1487M} for g modes.

\subsection{Revised asymptotic expansion}

\begin{figure}                 % Insertion d'une figure = objet flottant
  \includegraphics[width=9cm]{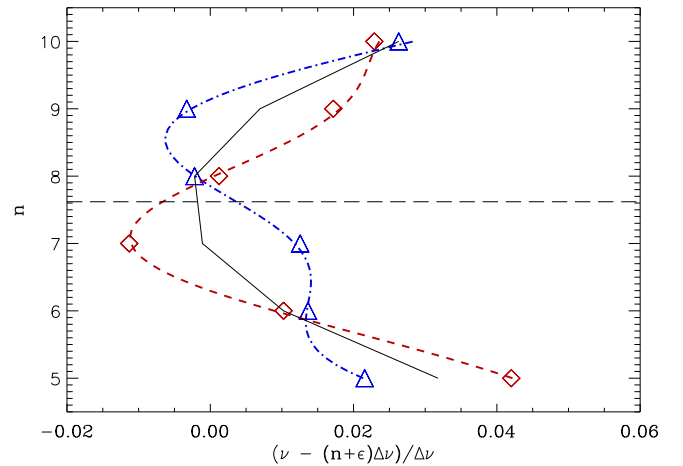}
  \caption{Synthetic \'echelle diagram of the radial modes of a typical red giant star with a large separation of $3.90$ $\mu$Hz. The glitch contributions are overplotted with a dashed red line and red diamonds for a typical clump star, and with a dot-dashed blue line and blue triangles for a typical RGB star. The horizontal dashed black line with a $n$ value of approximately 7.6 represents the $\nmax$ value. Only six radial modes are shown in this figure, which is typically the number of observed modes.}
  \label{fig:exemple_shift}
\end{figure}

We can now propose a complement to the universal pattern, which includes the modulation in order to investigate the consequence of the parameter variations in the observed shift. To achieve that, we used Eq. (\ref{dnu_local_lisse}), corresponding to the asymptotic universal pattern for radial modes, and added the glitch component modelled in Eq. (\ref{resultante}).

For radial modes, with $\nmax~=~\numax/\dnumoy~-~\varepsilon$, we get:

\begin{multline}
\hbox{\vrule width 12mm height 0mm}
\nu_{n} = \left(n + \varepsilon + \frac{\alpha}{2}(n-\nmax)^{2} \right.\\
\left. + \frac{\amp \per}{2\pi} \sin\left(\frac{2\pi(n-\nmax)}{\per} + \phase\right) \right)\dnumoy .
\label{univ_final}
\end{multline}
From \citet{1993A&A...274..595P}, we derive that the expression is in fact purely asymptotic, with the modulation term representing the signature of the structure discontinuity. The difference compared to the unmodulated pattern is shown in Fig. \ref{fig:exemple_shift}. As seen above, a local measurement of $\Dnu$ around $\numax$ will slightly overestimate the $\dnumoy$ value for clump stars and slightly underestimate it for RGB stars.

The identification of the modulation provides simultaneously a justification and an improvement of the asymptotic red giant universal oscillation pattern \citep{2011A&A...525L...9M}. Independent of this work, \citet{2013A&A...550A.126M} have shown that the $\epsilon (\Dnu)$ function observed in the red giant regime is the observational counterpart of the second-order asymptotic expansion. This work shows that the combination of the universal pattern and the modulation (Eq. \ref{univ_final}) provides a very precise determination of the eigenfrequency pattern with the parameters defined in Table 1.

\begin{table}[t]
\caption{Asymptotic parameters}\label{table-asymp}
\begin{tabular}{rll}
\hline
parameter & RGB & clump \\
\hline
$\varepsilon$ & \multicolumn{2}{c}{$0.601 + 0.632\log \dnumoy$} \\
$\nmax$ & \multicolumn{2}{c}{$\numax / \dnumoy - \varepsilon$}\\
$\alpha$ & \multicolumn{2}{c}{$0.015 \dnumoy^{-0.32}$}\\
$\amp$  & $0.06\dnumoy^{-0.88\pm0.05}$ & $0.07\dnumoy^{-0.74\pm0.05}$ \\
$\per$ & $3.08\pm0.65$ & $3.83\pm0.88$ \\
$\phase$ & $1.71\pm0.77$ & $-0.43\pm0.66$ \\

\hline
\end{tabular}

\scriptsize{with $\Dnu$ in $\mu$Hz.
}
\end{table}

\subsection{Curvature and evolutionary status}

Finally, we come back to the influence of the modulation on the apparent gradient of the radial frequency separation (Eq. \ref{dnu_local_curvature}). This gradient translates into an apparent curvature in the \'echelle diagram \citep{2013A&A...550A.126M}. Figure \ref{fig:exemple_shift} shows that the curvature of the radial ridge depends on the evolutionary status: around $\numax$, clump stars often show a higher curvature than RGB stars. This reproduces the observations (Fig. \ref{fig:curvature}) and justifies the use of a mean curvature term (Eq. \ref{alpha_measured}) used for removing the second-order effect. It demonstrates the relevance of the global picture provided by the universal pattern with glitches (Eq. \ref{univ_final}) for describing the red giant radial oscillation pattern.

\section{Conclusion\label{Conclusion.}}

In this study, we measured the frequency position of radial modes very precisely using a peakbagging method. From these frequencies, we calculated a mean value of $\Dnu$. Using the universal red giant oscillation pattern, we analysed the variation of $\Dnu$ as a function of the frequency. Modulations could be measured for half of the data set. We characterized the modulation in more than five hundred red giants as a glitch signature due to the second helium ionization zone. A value of the acoustic depth of this discontinuity has been extracted from this modulation and found consistent with predictions for RGB stars but not for clump stars. A different behaviour for the modulation was found in the phase parameter between the different evolution stages of the stars which leads to an interpretation of the results found by \citet{2012A&A...541A..51K}. The mass dependence of the phase of the modulation in clump stars needs some more theoretical work to be fully understood. Even if extracting any direct information on the helium content is certainly very difficult, this work opens the way for testing the second helium ionization zone in a large set of stars with ensemble asteroseismology. Finally, it shows that the global measurement of the large separation is less biases than the local measurement, hence provides better estimates of the stellar mass and radius.

\appendix

\section{Estimation of the uncertainties}\label{ap:1}

The $\chi^{2}$ method is used here to fit a model ($\ymodele$, $i = 1, N$) with a set of parameters $\textbf{a}$ over a range of data points $\ydonnees$ (which correspond to a combination of frequencies). It is written

\begin{equation}
   \chi^{2} = \sum_{i=1}^{N} \left(\frac{\ydonnees-\ymodele}{\erreur}\right)^{2} ,
   \label{khi2}
\end{equation}
with $\erreur$ the error on each data point.

The uncertainties on the parameters can be retrieved by the inversion of the Hessian matrix

\begin{equation}
   H_{kl} = \frac{\partial \chi^{2}}{\partial a_{k}\partial a_{l}} = 2\sum_{i=1}^{N} \left(\frac{1}{\erreur^{2}} \frac{\partial\ymodele}{\partial a_{k}} \frac{\partial\ymodele}{\partial a_{l}}\right) ,
   \label{hessian}
\end{equation}
when data are not correlated. The second-derivative terms were neglected.
\newline

Since the frequencies considered are correlated, the previous relations can not be valid. Another definition of Eq. (\ref{khi2}) must be used, which includes a covariance matrix taking the correlation between the parameters into account. Second derivative terms are usually neglected.

The covariance matrix indicates the correlation between the different data points. The data points are calculated using the set of radial mode frequencies. If we take two combinations of radial frequencies ($f$ and $g$):

\begin{equation}
   f = \sum_{n} u_{n}\nu_{n}
   \label{combinaison}
\end{equation}
\begin{equation}
   g = \sum_{n} v_{n}\nu_{n}
   \label{combinaison_bis}
\end{equation}
where $\nu_{n}$ are the radial frequencies, $u_{n}$ and $v_{n}$ are the coefficients used to calculate the large separation with the set of frequencies. For example, if $f$ correspond to the calculation of the local large separation $\Dnu(n) = (\nu_{n+1}-\nu_{n-1})/2$, we then have $u_{n-1} = -1/2$, $u_{n+1} = +1/2$ and the other coefficient $u_{n}$ equal to zero.

The elements of the covariance matrix \textbf{C} are then calculated as

\begin{equation}
   C_{ij} = \sum_{n} u_{n}v_{n}\sigma_{\ydonnees} \sigma_{y_{j}}
   \label{covariance}
\end{equation}
The elements of the covariance matrix are then set to zero when there is no correlation between the data points.

The $\chi^{2}$ calculation is then modified following
\begin{equation}
   \chi^{2} = (\textbf{y}-\textbf{y}(\textbf{a}))^{T} \textbf{W} (\textbf{y}-\textbf{y}(\textbf{a})) , 
   \label{khi2_corr}
\end{equation}
with $\textbf{W} = \textbf{C}^{-1}$ the inverse of the covariance matrix. If there is no correlation between the data points, we retrieve Eq. ({\ref{khi2}}). The Hessian matrix calculation has to take the correlation between the parameters into account as well:

\begin{multline}
   H_{kl} = \frac{\partial\chi^{2}}{\partial a_{k}\partial a_{l}} \\
 = 2\sum_{i=1}^{N} \sum_{j=1}^{N} W_{ij} \left(\frac{\partial y_{j}(\textbf{a})}{\partial a_{k}} \frac{\partial\ymodele}{\partial a_{l}}  + \frac{\partial\ymodele}{\partial a_{k}} \frac{\partial y_{j}(\textbf{a})}{\partial a_{l}}\right)
   \label{hessian_corr}
\end{multline}
Second-derivative terms were neglected like in Eq. (\ref{hessian})

The errors are then extracted from the eigenvalues of the inverse of this Hessian matrix.

\bibliographystyle{aa} % style aa.bst

\section*{Acknowledgements}

We acknowledge the entire Kepler team, whose efforts made these results
possible. Funding for this Discovery mission was provided by NASA's
Science Mission Directorate.

%
%%%%%%%% Meudon
MV, BM, CB, KB and RS acknowledge financial support from the ``Programme
National de
Physique Stellaire" (PNPS, INSU, France) of CNRS/INSU and from the ANR
program IDEE ``Interaction Des \'Etoiles et des Exoplan\`etes'' (Agence
Nationale de la Recherche, France).

%
%%%%%%%% Yvonne
YE acknowledges support from the UK Science and Technology
Facilities Council (STFC). Funding for the Stellar Astrophysics
Centre is provided by The Danish National Research Foundation
(Grant agreement no. DNRF106).

%
%%%%%%%% Thomas
TK acknowledges financial support from the Austrian Science Fund (FWF P23608).

%
%%%%%%%% Saskia
SH acknowledges financial support from the Netherlands Organisation
for Scientific Research
(NWO). The research leading to the presented results has received
funding from the European Research Council under the European
Community's Seventh Framework Programme (FP7/2007-2013) / ERC grant
agreement no 338251 (Stellar Ages).

%Sebastien Deheuvels

We thank S.Deheuvels for his indirect contribution to the estimate
of the uncertainties: his PhD manuscript is enlightening.

\listofobjects
\end{document}